\documentclass[a4paper]{article}

\usepackage{amsmath,amssymb,theorem}
\usepackage[english]{babel}

\newtheorem{teo}{Theorem}[section]
\newtheorem{lema}[teo]{Lemma}
\newtheorem{coro}[teo]{Corollary}

{\theorembodyfont{\rmfamily }
\newtheorem{defi}[teo]{Definition}
\newtheorem{nota}[teo]{Note}
\newtheorem{Remark}[teo]{Remark}}
\def\demo{\noindent \textit{Proof: }}

\begin{document}

\title{On the naturalness of Einstein's equation}

\author{Jos\'{e} Navarro, Juan B. Sancho} 

\footnotetext[1]{Department of Mathematics, University of Extremadura, Avenida de Elvas s/n, 06071, \\
Badajoz, Spain. \\
\textit{Email addresses:} navarrogarmendia@unex.es, jsancho@unex.es \\
The first author has been partially supported by a Spanish FPU grant, AP2006-02414}

\maketitle

\begin{abstract}
 We compute all 2-covariant tensors naturally constructed from a semiriemannian metric $g$ which are
 divergence-free and have weight greater than $ -2$.

 As a consequence, it follows a characterization of the Einstein tensor as the only,
 up to a constant factor, 2-covariant tensor naturally constructed from
 a semiriemannian metric which is divergence-free and has weight 0 (i.e., is independent of the unit of scale).
 Since these two conditions are also satisfied by the energy-momentum tensor of a relativistic space-time, we
 discuss in detail how these theorems lead to the field equation of General Relativity.
\end{abstract} \bigskip

\section{Introduction}

In General Relativity, it is supposed a field equation of the following type:
\begin{equation}\label{Fieldequation} G_2(g) \, = \, T_2
\end{equation} where $T_2$ is the energy-momentum tensor of the matter, $g$ is the Lorentz metric of space-time that
measures proper time and $G_2(g)$ is a suitable tensor constructed from $g$.

Since the energy-momentum tensor $T_2$ is symmetric and divergence-free, one is forced to choose for the left-hand side
of (\ref{Fieldequation}) a tensor $G_2(g)$ satisfying these two properties.

As it is well known, Einstein and Hilbert finally found in 1915 the so called Einstein tensor, $R_2(g) - \frac{1}{2} \,
r(g) g$, thus arriving to the field equation of the theory. This choice of $G_2(g)$ is suggested by a beautiful
classical result, first published by Vermeil (\cite{Vermeil}) and developed by Cartan (\cite{Cartan}) and Weyl
(\cite{Weyl}), which characterizes the Einstein tensor of a semiriemannian metric $g$ as the only, up to a constant
factor and \textit{the addition of a cosmological term} $\Lambda g$, 2-covariant symmetric and divergence-free tensor
whose coefficients are functions of the coefficients of the metric, its first and second derivatives and are linear
functions in these second derivatives. Since the very first days of the theory, this theorem has been one of the
cornerstones for the justification of the field equation of General Relativity.

Later on, this theorem was greatly improved by Lovelock (\cite{Lovelock}, \cite{LovelockDimension}) who proved that the
assumptions of symmetry and linearity on the second derivatives of the metric are superfluous in dimension $4$, which
is precisely the case of General Relativity. After that, there were many different attempts to improve and adapt this
theorem to other situations (see \cite{AndersonLovelock}, \cite{Anderson} and references therein).

In particular, Aldersley (\cite{Aldersley}) went further showing that if the geometric tensor $G_2(g)$ satisfies a
certain ``axiom of dimensional analysis$"$ then Lovelock's restriction to the second derivatives of the metric is also
superfluous (see Note \ref{NotaAld}).
\medskip

In this paper, we give a new characterization of the Einstein tensor. Recall that in General Relativity the time metric
$g$ measures proper time, so that changing the time unit entails replacing $g$ by $\lambda^2 g$; on the other hand, the
energy-momentum tensor is independent of the time unit. Therefore, we characterize the Einstein tensor of a
semiriemannian metric as the only 2-covariant tensor $G_2(g)$ intrinsically constructed from the metric $g$ which is
divergence-free and independent of the time unit; i.e., satisfies the condition:
\begin{equation} \label{Independencia}
G_2( \lambda ^2 g ) = G_2 (g) \quad , \quad \forall \ \lambda \in \mathbb{R}^+
\end{equation}

This characterization is valid in any dimension, there is no symmetry hypothesis and the dependence of the tensor
$G_2(g)$ is not even assumed to be through derivatives of the metric (see Section \ref{Statement}). It is surprising
how the apparently innocent property (\ref{Independencia}), never used before in order to characterize the Einstein
tensor, turns out to be much more restrictive than the divergence-free condition.

To be precise, we calculate in Theorem \ref{Resultado} all the 2-covariant tensors $G_2(g)$ naturally constructed from
a metric $g$ that are divergence free and have weight greater than $ -2$ (i.e., satisfy the condition $G_2(\lambda^2 g)
= \lambda^w G_2 (g) $, for $w > -2$). In a wider sense, this can be thought of as part of a more general programme
consisting in determining all the divergence-free tensors that can be constructed intrinsically from a semiriemannian
metric (see \cite{Anderson}, \cite{Annals}).

To obtain our result, we make use of the theory of natural bundles. Firstly, we explain what a ``tensor intrinsically
constructed from a metric$"$ is, reformulating the usual definitions (see \cite{Kolar}) in the most simple terms and
thus avoiding the categorical language of the standard treatment. Then we use a strong result of Stredder
(\cite{Stredder}) to prove Theorem \ref{Resultado}.

In Sections \ref{Discusion} and \ref{Newton}, we discuss in detail how our characterization of the Einstein tensor
improves the classical reasoning to derive Einstein's equation.

\section{Statement of the result} \label{Statement}

Let $X$ be a smooth  manifold of dimension $n \geq 2$.

We shall denote by $S^2_+T^*X \to X$ the fibre bundle of semiriemannian metrics with a given signature and
$\bigotimes^p T^*X \otimes \bigotimes^q TX \to X$ the vector bundle of $(p,q)$-tensors on $X$, whereas \textit{Metrics}
and \textit{Tensors} will stand for their sheaves of smooth sections, respectively. \medskip

To define what a ``tensor intrinsically constructed from a metric$"$ is, let us first consider a map:
$$ T \colon Metrics (X) \ \longrightarrow \ \mbox{\textit{Tensors}} (X) , \qquad
g \mapsto T(g)\ .$$ Then, let us suppose that this construction verifies certain physically reasonable
conditions:\medskip

1.- \textbf{Locality:} The value of the tensor $T(g)$ at any point $x$ only depends  on  the germ of $g$ at $x$.

Therefore, the map $T$ should be redefined as a morphism of sheaves:
$$ T \colon Metrics \ \longrightarrow \ \mbox{\textit{Tensors}}\ . $$

2.- \textbf{Regularity (differentiable dependence on the parameters):} If $\{g_s \}_{s\in S}$ is a family of metrics
depending smoothly on certain parameters, the family of tensors $\{ T(g_s)\}_{s\in S}$ also depends smoothly on those
parameters.

To be exact, let $S$ be a smooth manifold (the space of parameters) and let $U\subset X \times S$ be an open set. For
each ${s\in S}$, consider the open set in $X$ defined as $U_s := \{ x\in X \colon (x,s) \in U\}$. A family of metrics
$\{ g_s \in Metrics(U_s) \}_{s\in S}$ is said to be \textit{smooth} if the map $U \to S^2_+T^*X$, $(x,s) \mapsto
(g_s)_x$ is smooth. In the same way, a family of tensors $\,\{ T_s \in \mbox{\textit{Tensors}}(U_s)\}_{s\in S}\,$ is
said to be smooth if the map $\, U \to \bigotimes^p T^*X \otimes \bigotimes^q TX$, $(x,s) \mapsto (T_s)_x$ is smooth.

In these terms, the regularity condition express that for each smooth  manifold $S$, each open set $U \subset X \times
S$ and each smooth family of metrics $\{ g_s \in Metrics(U_s) \}_{s\in S}$, the family of tensors $\{ T(g_s) \in
\mbox{\textit{Tensors}}(U_s) \}_{s\in S}$ is smooth. \\

\goodbreak

3.- \textbf{Naturalness:} The morphism of sheaves $T$ is equivariant with respect to the action of local
diffeomorphisms of $X$.

That is, for each diffeomorphism $\tau \colon U \to V$ between open sets of $X$ and for each metric $g$ on $V$, the
following condition must be satisfied:
\begin{equation}\label{Naturalidad} T(\tau^* g) = \tau^* (T(g))\ . \end{equation}

Finally, it is also reasonable to consider homogeneity under changes of the unit of scale (see Section
\ref{Discusion}):

\begin{defi} A morphism of sheaves $T \colon Metrics \longrightarrow \textit{Tensors}$ is
said to be homogenous of \textbf{weight $w \in \mathbb{R}$} if it satisfies:
$$ T(\lambda^2 g) = \lambda ^w \, T(g) \, \qquad \quad \forall \, g , \   \forall \,
\lambda > 0 \ .$$ If the morphism $T$ has weight 0, it is said to be \textbf{independent of the unit of
scale}.\end{defi}

\begin{defi} Given a semiriemannian metric $g$ on $X$, tensors of the form $T(g)$, where
$T \colon Metrics \longrightarrow \textit{Tensors}$ is a regular and natural morphism of sheaves, are said to be
tensors \textbf{naturally constructed} from $g$ or \textbf{natural tensors} associated to $g$.

A tensor $T(g)$ naturally constructed from $g$ is homogenous of \textbf{weight} $w$ if the corresponding morphism of
sheaves $T$ is homogenous of weight $w$. \end{defi}

Notice that the above definition is quite general: a priori, the coefficients of a tensor $T(g)$ naturally constructed
from a metric $g$ are \textit{not} assumed to be functions of the coefficients of $g$ and their successive derivatives.

Given a semiriemannian metric $g$, we will denote its Ricci tensor by $R_2(g)$ and its scalar curvature by $r(g)$.

The main result of this paper is the following:

\begin{teo}\label{Resultado} Up to constant factors, the only divergence-free 2-covariant tensors of weight $w > -2$ naturally
constructed from a semiriemannian metric $g$ are the Einstein tensor $R_2(g) - \frac{r(g)}{2} \, g$, of weight 0, and
the metric itself $g$, of weight 2.
\end{teo}

\section{Normal tensors}

The normal tensors associated to a metric were first introduced by Thomas (\cite{Thomas}). Although scarcely used in
the literature, they present some important advantages when studying the space of jets of metrics at a point.

\begin{defi} Let $r \geq 1$ be a fixed integer. We denote by $\mathcal{N}^r (X)$ the
$\mathcal{C}^\infty(X)$-module of $(r+2)$-covariant tensors $T$ on $X$ having the following symmetries:

\begin{itemize}
\item[-] $T$ is symmetric in the first two and last $r$ indices:
$$ T_{ij k_1\cdots k_r} = T_{jik_1 \cdots k_r} \quad , \quad T_{ij k_1 \cdots k_r} = T_{ij k_{\sigma(1)} \cdots
k_{\sigma(r)}} \quad \forall \, \sigma \in S_r\ ,$$

\item[-] the symmetrization of $T$ over the last $r+1$ indices is zero:
$$ \sum_{\sigma \in S_{r+1}} T_{ik_{\sigma(1)} \cdots k_{\sigma(r+1)}} = 0\ .$$\end{itemize}
\noindent A tensor with these symmetries will be called a \textbf{normal tensor} of order $r$.\end{defi}

The space of normal tensors of order $r$ at a point $x \in X$ will be written $ \mathcal{N}^r_x \subset S^2T^*_xX
\otimes S^r T^*_xX$.

A simple computation shows that, in general, $\mathcal{N}^1_x = 0$.

To show how a semiriemannian metric $g$ produces a sequence of normal tensors $g^r \in \mathcal{N}^r(X)$, recall the
classical lemma:

\begin{lema}[Gauss] Let $y_1,\ldots , y_n$ be a system of normal coordinates at a point $x_0 \in X$ with respect
to the metric $g$. The coefficients $g_{ij}$ of the metric in these coordinates verify the system of equations:
\begin{equation}\label{Gausslema} \sum_{j=1}^n g_{ij} y_j = \pm y_i \qquad \quad  i = 1,\ldots , n
\end{equation}(the signs on the right side depend on the signature of the metric $g$).
\end{lema}

Let $y_1,\ldots , y_n$ be a system of normal coordinates at a point $x_0 \in X$ with respect to $g$ and let us denote:
$$ g_{ij,k_1 \cdots k_r} := \frac{\partial^r g_{ij}}{\partial y_{k_1} \cdots \partial y_{k_r}}
(x_0)\ . $$ It is clear that these coefficients $g_{ij,k_1 \cdots k_r}$ are symmetric in the first two and in the last
$r$ indices. Moreover, if we derive $r$ times the identity (\ref{Gausslema}) of the Gauss Lemma, we obtain:
$$ \sum_{\sigma \in S_{r+1}} g_{i\sigma(j), \sigma(k_1 ) \cdots \sigma (k_r) } = 0\ $$
so that the tensor

\begin{equation}
g^r_{x_0} := \sum_{ijk_1 \cdots k_r} g_{ij,k_1 \cdots k_r} \, \text{d}y_i \otimes \text{d}y_j \otimes \text{d}y_{k_1}
\otimes \cdots \otimes \text{d}y_{k_r} \end{equation} is a normal tensor of order $r$ at the point $x_0 \in X$ (notice
that it does not depend on the chosen coordinates).

\begin{defi} The tensor $g^r_{x_0}$ is called the $r$-th \textbf{normal tensor of the metric} $g$ at the point $x_0$.
\end{defi}

As a consequence of $\mathcal{N}^1 = 0$, the first normal tensor is always zero, $g^1 = 0$. However, the second normal
tensor $g^2$ is essentially equivalent to the Riemann-Christoffel tensor of $g$:
$$ R (D_1,D_2,D_3,D_4) := (\nabla_{D_1} \nabla_{D_2} D_4 - \nabla_{D_2}\nabla _{D_1}D_4 - \nabla_{[D_1, D_2]}D_4) \cdot
D_3\ ,$$ where $\cdot$ stands for the product with the metric $g$. A straightforward calculation in normal coordinates
proves the following:

\begin{lema}[\cite{Thomas}] The tensors $g^2 $ and $R$ are mutually determined by the
identities:
\begin{equation}\label{LemaThomas}
R_{ijkh} = g^2_{ihjk} - g^2_{ikjh} \quad , \quad g^2_{ijkh} = -(R_{ihjk} + R_{ikjh})/3\ .
\end{equation}
\end{lema}

Let $\mathcal{R}_{x_0}$ be the subspace of $\otimes^4 T^*_{x_0}X$ whose elements have the typical linear symmetries of
the Riemann-Christoffel tensors (i.e., they are skew-symmetric in the first two and last two indices and satisfy the
Bianchi linear identity).

The following isomorphism will be used later on:

\begin{coro}\label{Ordendos} The identities from (\ref{LemaThomas}) define a $Gl$-equivariant
linear isomorphism:
$$ \mathcal{N}^2_{x_0} \simeq \mathcal{R}_{x_0}\ .$$ \end{coro}

\begin{Remark} More generally, for each integer $r\geq 2$, the sequence $\{ g_x, g^2_x, g^3_x, \ldots, g^r_x \}$
of normal tensors of the metric $g$ at a point $x$ totally determines the sequence $\{ g_x, R_x,$ $ \nabla_xR,\ldots ,
\nabla^{r-2}_xR \}$ of covariant derivatives at a point of the Riemann-Christoffel tensor of $g$ and vice-versa (see
\cite{Thomas}).

The main advantage of using the normal tensors at a point is the possibility of expressing the symmetries of each
$g^r_x$ without using the other normal tensors, while the symmetries of $\nabla^r_x R$, for $r\geq 2$, depend on $R_x $
(recall the Ricci identities). \end{Remark}

\section{Tensors naturally constructed from a metric}

In order to determine all $(p,q)$-tensors of weight $w$ naturally constructed from a metric $g$, the Stredder-Slov\'{a}k
result, Theorem \ref{Stredder}, reduces the question to a problem of computing invariant tensors under the action of
the orthogonal group of the metric.

Fix a point $x_0 \in X$ and let $O := O(n^+ , n^-)$ be the orthogonal group of $(T_{x_0}X, g_{x_0})$, i.e., $O$ is the
group of isometries $\sigma \colon (T_{x_0}X, g_{x_0}) \to (T_{x_0}X , g_{x_0}) $ (recall that $g_{x_0}$ is not
necessarily positive definite). The space of all $(p,q)$-tensors $\bigotimes^p T^*_{x_0}X \otimes \bigotimes^q
T_{x_0}X$ and the symmetric powers $S^k\mathcal{N}^r_{x_0}$ are linear representations of $O$.

Given two linear representations $V,W$ of $O$,  let $\mathrm{Hom}_{O}(V,W)$ be the vector space of $O$-equivariant
$\mathbb{R}$-linear maps $V \to W$, and  let $V^{O}$ denote the subspace of $V$ whose elements are invariant under the
action of $O$.

\begin{teo}\label{Stredder} There exists an $\mathbb{R}$-linear isomorphism:
$$\begin{array}{c}
  \{ \mbox{(p,q)-Tensors of weight $w$ naturally constructed from $g$} \}  \\ \medskip
  \parallel \\ \medskip
   \bigoplus\limits_{\{ d_i \} } \mathrm{Hom}_{O} (S^{d_2}\mathcal{N}^2_{x_0} \otimes \cdots \otimes
S^{d_s}\mathcal{N}_{x_0}^s \ ,\  \bigotimes^p T^*_{x_0}X \otimes \bigotimes ^qT_{x_0}X )
\end{array}$$
where the summation is over all sequences of non-negative integers $\{d_2 , \ldots , d_s \} $, $s\geq 2$, satisfying
the equation:
\begin{equation}\label{Condicion}
2d_2 + \ldots + s\, d_s = p-q -w\ .
\end{equation}
If this equation has no solutions, the above vector space reduces to zero.
\end{teo}

\begin{Remark} This theorem essentially reformulates a result that can be found in Stredder
(\cite{Stredder}, Theorem 2.5). This author uses a more restrictive notion of ``natural tensor$"$, supposing that the
coefficients of a natural tensor $T(g)$ in a coordinate chart can be expressed as universal polynomials in the
coefficients of the metric, a finite number of its partial derivatives and the inverse of the determinant of the
metric. Moreover, he assumes a riemannian metric.

These restrictions were removed by Slov\'{a}k (\cite{Peetre}, \cite{SlovakJournal}, \cite{SlovakSemi}), the most important
tool being the non-linear Peetre theorem (\cite{Peetre}) stating that each local operator is $``$locally$"$ of finite
order.
\end{Remark}

\begin{Remark} If $\varphi \colon S^{d_2}\mathcal{N}_{x_0}^2 \otimes \cdots \otimes S^{d_s} \mathcal{N}_{x_0}^s \to \bigotimes^p
T^*_{x_0} X \otimes \bigotimes^q T_{x_0}X$ is an $O$-equivariant linear map, then the corresponding natural tensor
$T(g)$ is obtained by the formula:
$$ T(g)_{x_0} = \varphi \left( (g^2_{x_0} \otimes \stackrel{d_2}{\ldots} \otimes \, g^2_{x_0}) \otimes \cdots
\otimes (g^s_{x_0} \stackrel{d_s}{\ldots} \otimes \, g^s_{x_0}) \right) $$ where $(g^2_{x_0}, g^3_{x_0}, \ldots )$ is
the sequence of natural tensors of $g$ at the point $x_0 \in X$. \end{Remark}

\begin{Remark} \label{Par} The $O$-equivariant linear maps that appear in the theorem can be explicitly computed using the
isomorphism:
\begin{align*}
&\mathrm{Hom}_{O} \left( S^{d_2}\mathcal{N}_{x_0}^2 \otimes \cdots \otimes S^{d_s} \mathcal{N}_{x_0}^s \ ,\
\mbox{$\bigotimes^p$} T^*_{x_0}X \otimes \mbox{$\bigotimes^q$} T_{x_0}X \right) = \\
& \mathrm{Hom}_{O} \left( S^{d_2}\mathcal{N}_{x_0}^2 \otimes \cdots \otimes S^{d_s} \mathcal{N}_{x_0}^s \otimes
\mbox{$\bigotimes^p$} T_{x_0}X \otimes \mbox{$\bigotimes^q$} T^*_{x_0}X \ ,\ \mathbb{R} \right) \end{align*} and
applying the Main Theorem of the invariant theory for the orthogonal group $O$ (see \cite{Jaime} for a simple proof in
the semi-riemannian case).º

This theorem states that any $O$-equivariant linear map $S^{d_2} \mathcal{N}^2_{x_0} \otimes \cdots \otimes T^*_{x_0}X
\to \mathbb{R}$ is a linear combination of iterated contractions with respect to the metric $g_{x_0}$.

So, for a non zero linear map to exist, the total order (covariant plus contravariant order) of the space $S^{d_2}
\mathcal{N}^2_{x_0} \otimes \cdots \otimes T^*_{x_0}X $ has to be even.
\end{Remark}

\begin{coro}\label{Impar} The weight of an homogenous tensor naturally constructed from a metric is even.
\end{coro}

\demo Due to the previous Remark \ref{Par}, the total order (covariant plus contravariant order) of the space
$S^{d_2}\mathcal{N}^2_{x_0} \otimes \cdots \otimes S^{d_s}\mathcal{N}^s_{x_0} \otimes \bigotimes^p T_{x_0}X \otimes
\bigotimes ^q T^*_{x_0}X$ has to be even.

In other words, $d_2(2+2) + \cdots + d_s(2+s) + p +q$ is even, and so it is:
$$ 2d_2 + \cdots + sd_s + p + q \stackrel{(\ref{Condicion})}{=} (p-q-w) + p +q = 2p - w . $$  \hfill $\square$

\begin{coro}\label{Nulos} There are no $(p,q)$-tensors naturally constructed from a metric with
weight $w > p-q $ or $w= p-q -1$.
\end{coro}

\demo In these cases equation (\ref{Condicion}) has no solutions $\{ d_i \}$.  \hfill $\square$\bigskip

A remarkable characterization of the Levi-Civita connection, due to Epstein (\cite{Epstein}), follows from this
corollary:\smallskip

 {\it The only linear connection $\nabla (g)$ independent of the unit of scale
(i.e., $\nabla (\lambda^2g)=\nabla (g)$) which is naturally constructed from a semiriemannian metric $g$ is the
Levi-Civita connection.}\smallskip

Indeed, any other such linear connection $\overline\nabla (g)$ differs from the Levi-Civita connection $\nabla (g)$ in
a $(2,1)$-tensor of weight zero: $\, T(D_1,D_2):=\nabla_{D_1}D_2-\overline{\nabla}_{D_1}D_2$. By Corollary \ref{Nulos},
that tensor has to be zero.

\begin{coro}\label{Pesototalcero} There exists an $\mathbb{R}$-linear isomorphism:
$$\begin{array}{c}
  \{ (p,q)\mbox{-Tensors of weight $w=p-q$ naturally constructed from $g$} \} \\
   \medskip   \parallel \\ \medskip
  (\bigotimes^p T^*_{x_0} X \otimes \bigotimes ^q T_{x_0} X)^{O}
\end{array}$$
\end{coro}

\demo If $w = p-q$, then equation (\ref{Condicion}) only has the trivial solution $\{ d_i = 0\}$, so in this case the
space of tensors under consideration is isomorphic to:
$$ \mathrm{Hom}_{O} (\mathbb{R}\ ,\ \mbox{$\bigotimes^p$} T^*_{x_0} X \otimes
\mbox{$\bigotimes^q$} T_{x_0}X) = (\mbox{$\bigotimes^p$} T^*_{x_0}X \otimes \mbox{$\bigotimes^q$} T_{x_0}X ) ^{O}\ .$$
\hfill $\square$

\begin{coro} \label{CasoEinstein} There exists an $\mathbb{R}$-linear isomorphism:
$$\begin{array}{c}
  \{ (p,q)\mbox{-Tensors of weight $w=p-q-2$ naturally constructed from $g$} \} \\
  \medskip   \parallel \\ \medskip
  \mathrm{Hom}_{O} (\mathcal{R}_{x_0}\ ,\  \bigotimes^p T^*_{x_0} X \otimes
\bigotimes^q T_{x_0}X )
\end{array}$$
\end{coro}

\demo If $w= p-q-2$, then equation (\ref{Condicion}) has the only solution $ d_2 = 1, d_3 = d_4 = \cdots = 0 $, thus,
in this case the space of tensors under consideration is isomorphic to:
$$ \mathrm{Hom}_{O} (\mathcal{N}^2_{x_0}\ ,\ \mbox{$\bigotimes^p$} T^*_{x_0} X \otimes
\mbox{$\bigotimes^q$}T_{x_0}X ) \stackrel{(\ref{Ordendos})}{=} \mathrm{Hom}_{O}(\mathcal{R}_{x_0}\ ,\
\mbox{$\bigotimes^p$} T^*_{x_0}X \otimes \mbox{$\bigotimes^q$} T_{x_0}X)$$ \hfill $\square$

\begin{Remark} If $\,\varphi \colon \mathcal{R}_{x_0} \to \bigotimes^p T^*_{x_0}X \otimes \bigotimes^q T_{x_0}X\,$ is an
$O$-equivariant linear map, then the corresponding tensor $T(g)$ naturally associated to $g$ is obtained by the
formula:
$$ T(g)_{x_0} = \varphi (R(g)_{x_0})$$ where $R(g)_{x_0}$ is the Riemann-Christoffel tensor of $g$ at the point $x_0$.
\end{Remark}

\section{The Einstein tensor}

Recall that $R_2(g)$ stands for the Ricci tensor of the metric $g$ and $r(g)$ for its scalar curvature.

\begin{teo}\label{LemaEinstein} Any $2$-covariant tensor of weight $0$ naturally
constructed from a me\-tric $g$ is an $\mathbb{R}$-linear combination of $R_2(g)$ and $r(g)\, g$.
\end{teo}

\demo By Corollary \ref{CasoEinstein}, the space of tensors under consideration is isomorphic to:
$$ \mathrm{Hom}_{O} (\mathcal{R}_{x_0}\ ,\ T^*_{x_0}X \otimes T^*_{x_0} X ) =
\mathrm{Hom}_{O} (\mathcal{R}_{x_0} \otimes T_{x_0}X \otimes T_{x_0}X\ ,\ \mathbb{R})\ . $$ The latter vector space is,
according to the Main Theorem for the orthogonal group, generated by the operators of iterated contractions. Due to the
symmetries of the elements of $\mathcal{R}_{x_0}$, these generators reduce (up to signs) to the following two:
$$ C_{13,24,56} \quad \mbox{and} \quad C_{13,25,46}$$ where each pair of indices denotes the contraction of these
indices.

These two operators correspond with the maps $\mathcal{R}_{x_0} \to T^*_{x_0} X \otimes T^*_{x_0} X$ defined by:
$$ R \mapsto C_{13,24}(R) \, g_{x_0} \qquad \mbox{and} \qquad R \mapsto C_{13}(R)\ .$$
The first one produces the tensor $r(g) \, g$ and the second one $R_2(g)$. \hfill $\square$ \bigskip

\noindent \textit{Proof of Theorem \ref{Resultado}:} By Corollary \ref{Impar} and Corollary \ref{Nulos}, there are no
such tensors with a weight different from $w = 0,2$.

If $w = 0$, there only exists the Einstein tensor, as follows from Theorem \ref{LemaEinstein}, together with the well
known identities:
$$ \mathrm{div}_g R_2(g) = \frac{1}{2}\, \mathrm{grad}_g r(g) \quad , \quad \mathrm{div}_g r(g) g = \mathrm{grad}_g
r(g)\ .$$

If $w=2$, the only 2-covariant tensor naturally constructed from $g$ is the metric itself $g$ (as follows from
Corollary \ref{Pesototalcero} and $  (\otimes^2 T^*_{x_0} X )^{O} \ = \  < g_{x_0}> \, )$, which is divergence-free.
\hfill $\square$

\begin{nota}\label{NotaAld} The case $w=0$ in Theorem \ref{Resultado} is closely related to a result of
Aldersley (\cite{Aldersley}). This author considers a divergence-free 2-contravariant tensor $A^2$ constructed from a
metric $g$, whose coefficients $A^{ij}$ depend on a finite number of derivatives of the coefficients of the metric. It
is also assumed that these coefficients satisfy, with respect to a suitable system of coordinates, the following
condition (that he calls \textit{axiom of dimensional analysis}):
\begin{equation*}
A^{ij} ( g_{rs} , \lambda g_{rs,t_1} , \lambda^2 g_{rs,t_1t_2} , \ldots , \lambda^k g_{rs,t_1 \cdots t_k} ) = \lambda^2
A^{ij} (g_{rs} , g_{rs,t_1}  , \ldots , g_{rs,t_1 \cdots t_k} )
\end{equation*}
for all $\lambda > 0$. Then it is proved that $A^2$ coincides (up to a constant factor) with the contravariant Einstein
tensor $G^2$.

Although the above axiom is not intrinsic, it is not difficult to show that Aldersley's axiom for $A^2$ is equivalent,
in the case of a natural tensor, to the condition of $A^2$ having weight $-4$ or, in other words, of $A_2$ having
weight 0, where $A_2$ is the 2-covariant tensor metrically equivalent to $A^2$. \end{nota}

\section{Einstein's equation}
\label{Discusion}

In the theory of General Relativity, space-time is a differentiable manifold $X$ of dimension 4 endowed with a Lorentz
metric $g$, i.e., a semiriemannian metric of signature $(+,-,-,-)$, called the time metric. The proper time of a
particle following a trajectory in $X$ is defined to be the length of that curve using the metric $g$. So that if the
metric $g$ is changed by a proportional one $\lambda^2 g$, with $\lambda \in \mathbb{R}^+$, then the proper time of
particles is multiplied by the factor $\lambda$. Therefore, replacing the metric $g$ by $\lambda^2g$ amounts to a
change in the time unit. It is a convention to define the relativistic space-time as a pair  $(X,g)$, but it would be
more accurate to think of it as a pair $(X, \{ \lambda^2 g\})$, bearing in mind that fixing a time unit is the physical
counterpart of choosing a metric in the family $\{\lambda^2 g \}$.

As we have said before, the mass-energy distribution of space-time is represented by means of a 2-covariant tensor
$T_2$, known as the \textit{energy-momentum tensor}. For each infinitesimal orthonormal frame $\{ \partial_t,
\partial_{x_1}, \partial_{x_2} , \partial_{x_3} \} $ at a point $p \in X$, the
scalar $T_2(\partial_t, \partial_t)$ is the mass-energy density at the point $p$ measured in that frame.

Two properties of this energy-momentum tensor are essential for our discussion. The first one is the assumption that
the mass-energy distribution satisfies, infinitesimally, a \textit{conservation principle}, stated by the condition:
$$ \mathrm{div}\, T_2 = 0$$

The other fundamental property deals with its dimensional analysis. The tensor $T_2$ is of dimension $ L^{-3} T^2 M^1$;
that is, if we change the units of length, time and mass in such a way that the length of each arc is multiplied by
$l$, the duration of each time interval is multiplied by $t$ and the mass of each object is multiplied by $m$, then the
energy-momentum tensor in these new units is the old one multiplied by the factor $l^{-3}t^2m$.

Recall that if we measure the mass of an object by the gravitational acceleration that it produces, i.e., we fix the
mass unit to be the mass that, at the distance of one unit, produces a gravitational acceleration of one unit, we can
reduce the mass unit to the units of length and time, which can still be fixed arbitrarily.

If we do so in newtonian gravitation, where the gravitational acceleration is proportional to the mass and inversely
proportional to the square of the distance, then mass is of dimension $L^3 T^{-2}$. Therefore, the mass-momentum tensor
of the newtonian theory is of dimension $L^0 T^0$, i.e., it does not depend on the fixed units.

In General Relativity, by analogy with the corresponding tensor in the newtonian case, the energy-momentum tensor $T_2$
of space-time is therefore assumed to be \textit{independent of the time unit}, i.e., $T_2$ remains the same when we
replace $g$ by $\lambda ^2g$, with $\lambda \in \mathbb{R}^+$.

To sum up, the energy-momentum tensor $T_2$ of a relativistic space-time is a divergence-free tensor independent of the
time unit.
\medskip

As for the field equation, in General Relativity gravitation is understood to be a manifestation of the geometry of
space-time; the way celestial bodies move is not explained by means of a force, but by the curvature of space-time. In
the newtonian theory, the newtonian potential determines, via the Poisson equation, the mass distribution. In the
relativistic theory, the newtonian potential is replaced by the space-time metric $g$, so that we would expect this
metric $g$ to determine the mass-energy distribution. Hence, the energy-momentum tensor $T_2$ should be equal to some
tensor $G_2(g)$ intrinsically constructed from the metric $g$:
\begin{equation}\label{FieldEquation} G_2(g) = T_2 \end{equation}
As we have already said, the energy-momentum tensor $T_2$ is divergence-free and does not depend on the time unit, so
the ``geometric$"$ tensor $G_2(g)$ also has to fulfill this two properties:
$$ \mathrm{div}\, G_2(g) = 0 \qquad \mbox{and} \qquad G_2 (\lambda^2g) = G_2(g)
\quad \forall \, \lambda \in \mathbb{R}^+$$ Then, Theorem \ref{Resultado} states that $G_2(g)$ has to be proportional
to the Einstein tensor $R_2(g) - \frac{1}{2}\, r(g) g$ and, therefore, the field equation (\ref{FieldEquation}) has to
be necessarily Einstein's one:
$$ R_2(g) - \frac{r(g)}{2} \, g = \alpha \, T_2 $$ for some constant $\alpha \in \mathbb{R}$.\medskip

\noindent \textbf{Remark on the cosmological constant:} Let us briefly remark that our theorem does not discard the
existence of a cosmological term in the field equation. It refines the geometric construction for the left-hand side of
the field equation and, therefore, it only suggests that the cosmological term lives, if it exists, on the right-hand
side of the equation.

\section{Newton's Inverse-Square Force Law}
\label{Newton}

This derivation of Einstein's equation has used the validity in the limit of Newton's Law of Universal Gravitation.
Recall that the energy-momentum tensor of a relativistic space-time is independent of the time unit, by analogy with
the newtonian theory, where the mass-momentum tensor is independent of the units of scale. As we said before, this
independence is due to the fact that the newtonian gravitational acceleration is proportional to the mass and inversely
proportional to the square of the distance (Newton's Law).

But a slight variation in the argument allows one to derive Einstein's equation from the single assumption of a
gravitational acceleration that \textit{goes to zero at infinity}, thus avoiding the famous Inverse-Square Force Law.

To see this, let us suppose a ``newtonian$"$ theory based on a Law of Universal Gravitation of the form:
\begin{equation}\label{LGU}
F = G \, m  m'  f(r)
\end{equation}
where $F$ is the gravitational force, $G \in \mathbb{R}$ is the universal constant, $m$ and $m'$ are the masses under
consideration, $r$ is the distance between them and $f(r)$ is a continuous function on the distance that goes to zero
at infinity.

If we change the length unit in such a way that distances are multiplied by $\lambda$, then, as the universal constant
$G$ and the force have some dimension, there exists a continuous function $h (\lambda)$, such that:
$$ F = h(\lambda )\, G m m' f(\lambda r) $$
Using both equations, we get: $$ f(r) = h(\lambda ) f(\lambda r) $$ and taking $r=1$, we obtain that, for some constant
$c := f(1)^{-1} \in \mathbb{R}$:
$$ f(\lambda r ) = c \, f(\lambda) f(r) $$ So
that $c \, f \colon (\mathbb{R}^+ , \cdot ) \to (\mathbb{R}^+ , \cdot ) $ is a continuous homomorphism and an easy
exercise shows that $f$ is in fact a monomial:
$$ f (r)  = \frac{ a }{ r^b}$$
where $a \in \mathbb{R}$ is constant and $b \in \mathbb{R}^+$ because $f$ goes to zero as $r$ goes to infinity.

Therefore, the gravitational acceleration has to be proportional to the mass and inversely proportional to some
positive power $b \in \mathbb{R}^+$ of the distance.

In this ``newtonian$"$ theory, mass would be of dimension $L^{1+b}T^{-2}$ and the mass-momentum tensor would then be of
dimension $L^{b-2}T^0$. For the field equation of the relativistic version of this theory, we should look for a tensor
$G_2(g)$ of weight $w=b-2 > -2$. But Theorem \ref{Resultado} states that there are no tensors $G_2(g)$ with such a
weight $w$, except for the cases $b = 2$ of the Einstein tensor and $b = 4$, that would produce a field equation $g
=\alpha \, T_2$ ($\alpha \in \mathbb{R}$) which is physically absurd.

It is certainly satisfying that the mere requirement of the existence of a field equation as (\ref{FieldEquation})
(together with the fact that the gravitational acceleration goes to zero with distance) has such impressive
consequences.

\bigskip
\small{ \noindent \textbf{Acknowledgements:} We would like to thank R. Faro and C. Tejero for their helpful comments
and the invaluable collaboration of Prof. J. A. Navarro.}

\end{document}